\documentclass[12pt,psfig]{article}
\usepackage{amssymb}
\usepackage{amsmath,amsthm}
\usepackage{amsfonts}
\usepackage{graphicx}
\usepackage{graphics}
\usepackage{indentfirst}
\usepackage{color,graphicx}
\usepackage{caption}
\usepackage[utf8]{inputenc} 
\usepackage[english]{babel}
\usepackage[nottoc]{tocbibind}
\usepackage[usenames,dvipsnames]{xcolor}
\usepackage[symbol]{footmisc}
\numberwithin{equation}{section}

\usepackage{hyperref}

\begin{document}
%\title{ui}\maketitle
\begin{center}\Large\textbf{$F(\phi)T$-Gravity
and Inflationary Natural Model}
\end{center}
\vspace{0.75cm}
\begin{center}{\large Feyzollah Younesizadeh and \large Davoud
Kamani} {\footnote{\textcolor{Magenta}
{Corresponding author}}}
\end{center}
\begin{center}
\textsl{\small{Department of Physics, Amirkabir University of
Technology (Tehran Polytechnic) \\
P.O.Box: 15875-4413, Tehran, Iran \\
e-mails: fyounesizadeh@aut.ac.ir , kamani@aut.ac.ir \\}}
\end{center}
\vspace{0.5cm}

%%%%%%%%%%%%%%%%%%%%%%%%%%%%%%%%%%%%%%%%%%%%%
\begin{abstract}

By applying a particular kind of modified gravity, we 
study the inflation. Precisely, we extend our investigations
beyond the Einstein's gravity to explore the Natural
inflation model via the term
$F(\phi)T$. We compute the inflation dynamics to
derive the slow-roll parameters, i.e., the
tensor-to-scalar ratio ``$r$'' and the
scalar spectral index ``$n_s$''. This modified form of the
gravity yields not only the 
predictions of the original models but
also better-fitting with the Planck/BICEP/Keck data.

\end{abstract}


%$$$$$$$$$$$$$$$$$$$$$$$$$$$$$$$$$$$$$$$$$$$$$$$$$$$$$$$
\newpage
\section{Introduction}

The inflation theories (\textcolor{blue}{Guth .1981; Starobinsky
.1980; Albrecht \& Steinhardt .1982}),
characterize the early universe as an
exponentially accelerating phase. They offer
a Natural solution to the horizon and flatness
problems of the universe, induced by the Big-Bang model.
The inflation that takes place close to
a (local) maximum of the potential, surely satisfies
the flatness conditions.
Besides, the conventional model of the current cosmology,
also known as the ``hot Big-Bang model'' (\textcolor{blue}{Linde
.1982; Guth .1981; Martin .2004}),
is partly based on the
inflation hypothesis. This phase of the accelerated expansion
occurred at very high energy in the very early cosmos. It is
defined as falling between $10^{15}$ GeV and the Big-Bang
nucleosynthesis.

According to the observations of the cosmic
microwave background (CMB), the
inflation has been tested with remarkable precision.
The most recent CMB data from the (\textcolor{blue}{BICEP,
Keck collaboration .2021; Planck Collaboration .2020})
and the Planck satellite 
(\textcolor{blue}{G. Hinshaw et al. [WMAP
Collaboration] .2013; Planck Collaboration .2014})
prominently put strong constraints 
on the inflationary theories, particularly
on the tensor-to-scalar ratio ``$r$'' 
and the spectral index ``$n_s$''.
According to the observable ``$r$'', some of the single-field
inflationary models have been accepted
and others, such as the chaotic model, have been ruled out.

The single-field models are the straightforward road
to studying the inflationary cosmology.
Besides, these models are the simplest models.
Among the well-known inflationary models with
a single monomial potential, is the power-law
potential (\textcolor{blue}{Linde .1983}). 
In this model, the inflation
takes place at the high values of the inflaton field $\phi$.
Besides, there are some models, motivated by the particle
physics, with nearly flat potentials.
In addition, in the hilltop models (\textcolor{blue}{Boubekeur
\& Lyth .2005; Linde .1982; Kinney \& Mahanthappa .1996})
the inflaton rolls away from an unstable equilibrium.
Furthermore, the D-brane inflation (\textcolor{blue}{Kachru
et al .2003; Burgess et al .2001; 
Garcia-Bellido et al .2002; Martin
et al .2014; Kallosh et al .2019})
suggests that the inflation may come from other dimensions.
Finally, the natural inflation 
(\textcolor{blue}{Freese et al .1990;
Goncharov \& Linde .1984; Dvali \& 
Tye .1999; Kallosh et al .2013;
Carrasco et al .2015; Kallosh \& 
Linde .2015; Kallosh \& Linde .2019;
Galante et al .2015; Adams et al .1993}),
which is based on the periodic potential, is a well-known
example in the inflationary cosmology.

This inflation is elaborated via a Pseudo-Nambu-Goldstone
boson with a nearly flat potential.
The other inflationary models are: the exponential
tails (\textcolor{blue}{Goncharov \& Linde .1984; Dvali \& Tye 
.1999}) in the string theory and supergravity,
cosmological attractors (\textcolor{blue}
{Kallosh \& Linde .2021}), $\alpha$-attractors
(\textcolor{blue}{Kallosh et al .2019; Kallosh
et al .2013; Carrasco et al .2015; 
Galante et al .2015}) through the supergravity
and conformal symmetry, canonical and
non-canonical inflation (\textcolor{blue}{Mishra et al .2022}), and the well-known
$R^2$-inflationary model that was initiated
by Starobinsky (\textcolor{blue}
{Starobinsky .1980; Whitt .1984; Younesizadeh
\& Rezaie .2022}).

An important approach to the inflation 
is an extension of the GR action
via the geometric components, the 
matter components, or both of them.
Therefore, the modified gravities 
with $f(R)$, $f(T)$, $f(G)$,
$f(R, T)$ and so on, have been 
appeared (\textcolor{blue}{Buchdahl
.1970; Flanagan .2004; Ferraris et 
al .1994; Bengochea \& Ferraro
.2009; Nojiri \& Odintsov .2005; 
Harko et al .2011; Ferraro \&
Fiorini .2007; Nojiri \& 
Odintsov .(2004 , 2005, 2006); Allemandi
et al .2005; Bertolami et al .2007}).
The modified gravity theories play essential role
in different parts of the cosmology. 

Among the modified gravities, the $F(\phi)T$-gravity
(\textcolor{blue}{Dzhunushaliev et al .2014;
Liu et al .2016; Chen \& Kung .2018;
X. Zhang et al .2022}) 
possesses the non-minimal coupling of the 
inflaton field $\phi$ with ``$T$'', i.e. the trace of
energy-momentum tensor. 
This term is generally motivated
by the quantum gravity. It gives
a description of the unified gravity
with other fields (\textcolor{blue}{Dzhunushaliev et al .2014;
Liu et al .2016; Chen \& Kung .2022; Rubio \& Wetterich .2017}). 
The $F(\phi) T$-gravity is
an extension of the $f(R,T)$-gravity. 
In Ref. (\textcolor{blue}{X. Zhang et al .2022}) 
the slow-roll inflation 
in the presence of the term
$F(\phi)T=\sqrt{\kappa}\;\phi T$ has been investigated.
It revealed that the Chaotic, Natural
and Starobinsky models behave in better agreement
with the observational data. 

In this paper, we introduce a 
modified gravity with the Lagrangian density
${R}/{2\kappa}+\beta F(\phi)T+\mathcal{L}_{m}(\phi)$.
The only restriction is that $F(\phi)$ obviously vanishes
when the inflaton field $\phi$ decays.
At first, we work out with an arbitrary functional
$F(\phi)$ and the scalar potential $V(\phi)$.
Then, special forms of $F(\phi)$ 
and $V(\phi)$ will be considered.

The structure of the paper is as follows. In Sec. \ref{200},
we briefly review the single-field 
inflation in the Einstein gravity.
In Sec. \ref{300}, the inflation due to the term
$F(\phi)T$ and a general potential $V(\phi)$
will be computed. We show the influence of the term $F(\phi)T$
on the natural inflationary model and extract the
observables and the slow-roll parameters.
In Sec. \ref{400}, for special forms of the functionals
$F(\phi)$ and $V(\phi)$, we present a comparative analysis
between the calculated data and the available data
by the Planck 2018. Section \ref{500} 
is devoted to the conclusions.

%%%%%%%%%%%%%%%%%%%%%%%%%%%%%%%%%%%%%%%%%%%%%%%%%%%%%%%%
\section{A brief review of the inflation in the Einstein gravity}
\label{200}

In the framework of the Einstein's gravity, we first give
a quick overview of the single-field
inflationary model. The action in the general relativity
consists of a matter part and the Einstein-Hilbert term
\begin{equation}
S_{\rm GR}=\int{{\rm d}^4x\sqrt{-g}}\left(\frac{R}{2\kappa}+
\mathcal{L}_{m}\right),
\end{equation}
where, in the Natural units $c=1$, and hence
$\kappa=8\pi G$. This action gives the Einstein's equation
$R_{\mu \nu}-\frac{1}{2} g_{\mu \nu}R =\kappa T_{\mu \nu}$
for the evolution of the space-time metric. The tensor
$T_{\mu\nu}$ represents the energy-momentum tensor,
extracted from $\mathcal{L}_{m}$.

For the perfect fluid, the energy-momentum tensor is
$T^\mu _{\;\;\;\nu}= {\rm diag}\{-\rho ,p,p,p\}$,
where $\rho$ and $p$ stand for the energy density
and pressure, respectively.
We apply the metric signature as $(-,+,+,+)$.
Combining the Friedmann-Robertson-Walker (FRW)
metric with the Einstein equation yields
the two Friedmann equations for the background metric evolution
\begin{equation}
\label{2}
H^2=\frac{\kappa \rho}{3},\qquad \frac{\ddot{a}}{a}=
-\frac{\kappa}{6}(3p+\rho),
\end{equation}
where the Hubble function is $H\equiv{\dot a}/{a}$, and $a=a(t)$
stands for the scale factor.
The continuity equation for the system is
\begin{equation}
\label{4}
\dot{\rho}+3H(\rho+p)=0.
\end{equation}
Besides, we have $\dot{H}= -\frac{\kappa}{2}(\rho+p)$.

During the adequate inflationary phase (the slow-roll
inflation with the last $50-60$ e-folds), the single
scalar field $\phi$ manifestly possesses a dominant effect.
Its contribution to the matter Lagrangian is given by
\begin{equation}
\label{5}
\mathcal{L}_{m}=-\frac{1}{2}g^{\mu \nu}
\partial_{\mu}\phi
\partial_{\nu}\phi-V(\phi)=\frac{1}{2}
\dot \phi^2-V(\phi),
\end{equation}
where it is assumed that the inflaton field
is homogeneous in space and merely
depends on the cosmic time ``$t$''. The
energy-momentum tensor, corresponding to this
field has the feature
\begin{equation}
T_{\mu \nu}=\partial_{\mu}\phi\partial_{\nu}\phi
+g_{\mu \nu}\left(\frac{1}{2} \dot \phi^2
-V(\phi)\right).
\end{equation}

For the latter purposes, the trace of this tensor will be
needed, which is $T=g^{\mu\nu}T_{\mu\nu}=\dot \phi^2-4V(\phi)$.
When an inflaton field is spatially
homogeneous, its energy-momentum
tensor reliably resembles a perfect fluid one. The associated
energy density ``$\rho$'' and the pressure ``$p$'' are
represented by the elements $T_{00}
=\frac{\dot \phi^2}{2}+V(\phi)$ and
$T_{ij}=\left(\frac{\dot \phi^2}{2}-
V(\phi)\right) g_{ij}$, respectively.
Thus, the continuity equation of the scalar field \eqref{4}
finds the form
\begin{equation}
\ddot{\phi}+3H\dot \phi+V_{,\phi}=0.
\end{equation}

In addition, the Friedman equations $H^2=\frac{\kappa}{3}
\left(\frac{\dot \phi^2}{2}+ V(\phi)\right)$ and
$\frac{\ddot{a}}{a}=-\frac{\kappa}{3} \big(\dot
\phi^2-V(\phi)\big)$ are associated with the inflation dynamics.

The conventional slow-roll inflation
model drastically requires the conditions
$\frac{\dot \phi^2}{2}\ll V(\phi)$
and $\vert\ddot \phi \vert\ll \vert 3H \dot \phi\vert$.
It clarifies that for occurring the inflation
the potential energy of the scalar
field should exceed the total energy of the
universe. In the slow-roll approximation,
the equations for the scalar field and
the background metric finds the features
\begin{equation}
3H\dot \phi+V_{,\phi}\simeq0,
\end{equation}
\begin{equation}
H^2\simeq \frac{\kappa}{3}V(\phi).
\end{equation}

In the inflationary theory, a lot of researches begin with
the slow-roll approximation, which is dependent
on a variety of factors. This approximation is typically
specified by the smallness of the slow-roll parameters
(\textcolor{blue}{Liddle \& Lyth .2000}),
\begin{equation}
\label{14}
\epsilon_{\rm E}=\frac{1}{16\pi G}\left(
\frac{V_{,\phi}}{V}\right)^2\ll 1, \qquad
\eta_{\rm E}=\frac{1}{8\pi G}\frac{V_{,\phi \phi}}{V}\ll 1,
\end{equation}
where the subscribe ``E'' indicates the Einstein's gravity.

To obtain the field value at the horizon crossing, one must
accurately count the backward
from the end of the slow-roll and 
ensure that at least 50 e-folds
are formed. The field value at the horizon
crossing and also the corresponding slow-roll parameters
give the following expressions for the spectral index and
the tensor-to-scalar ratio
\begin{equation}
\label{2.10}
n_{\rm s}=1+2\eta_{\rm E} -6 \epsilon_{\rm E} ,
\qquad r= 16\epsilon_{\rm E}.
\end{equation}
These quantities extremely are 
used to compare the data (\textcolor{blue}{Hwang \& Noh .2001}).

%%%%%%%%%%%%%%%%%%%%%%%%%%%%%%%%%%%%%%%%%%%%%%%%%%%%
\section{The inflation due to the $F(\phi)T-$gravity:
general calculations}
\label{300}

In our modified gravity model, we 
incorporate the extra term $F(\phi)T$.
It consists of the energy-momentum tensor trace
and an arbitrary functional of the inflaton
\begin{equation}
\label{16}
S=\int{{\rm d}^4x\sqrt{-g}}\left(\frac{R}{2\kappa}
+\beta F(\phi)\;T+\mathcal{L}_{m}\right).
\end{equation}
where ``$\beta$'' is a constant parameter. 
We shall demonstrate that
for the small values of ``$\beta$'',
the inflationary model predictions
are extremely fitted with the observations. After
the inflation, when the inflaton decays, our model
returns to the Einstein's gravity, as expected. In this case,
there should be no concern
about the underlying paradoxes or unintended consequences.

The action \eqref{16} conveniently 
defines the Einstein equation
with an effective energy-momentum tensor
\begin{equation}
\label{17}
R_{\mu \nu}-\frac{1}{2}g_{\mu \nu}R =
\kappa T_{\mu \nu}^{({\rm eff})},
\end{equation}
The relation between the trace $T$ and the general
functional $F(\phi)$ is also 
characterized as in the following
\begin{equation}
\label{7.1}
T_{\mu \nu}^{({\rm eff})}=\frac{-2}
{\sqrt{-g}}\frac{\partial(\sqrt{-g}
\mathcal{L}^{({\rm eff})}_{m})}{\partial g^{\mu\nu}},
\end{equation}
where $\mathcal{L}^{({\rm eff})}_{m}$ is a
combination of the matter Lagrangian density along with the
additional term $F(\phi)T$,
\begin{equation}
\mathcal{L}^{({\rm eff})}_{m}=\beta F(\phi)T+\mathcal{L}_{m}.
\end{equation}
Consequently, the tensor \( T_{\mu \nu}^{({\rm eff})} \), as
specified in Eq.\eqref{7.1}, takes the feature 
\begin{equation}
\label{24}
T_{\mu \nu}^{({\rm eff})} = T_{\mu \nu}-2\beta
F\left(T_{\mu \nu}
-\frac{1}{2}Tg_{\mu \nu}+\Theta_{\mu \nu}\right),
\end{equation}
\begin{equation}
\Theta_{\mu \nu}\equiv g^{\alpha \beta} 
\frac{\delta T_{\alpha
\beta}}{\delta g^{\mu \nu}}=-2T_{\mu \nu}+g_{\mu \nu}
\mathcal{L}_{m}-2g^{\alpha \beta}
\frac{\delta^2 \mathcal{L}_{m}}{\delta
g^{\mu \nu}\delta g^{\alpha \beta}}.
\end{equation}
Note that $T_{\mu \nu}=\frac{-2}{\sqrt{-g}}
\frac{\partial(\sqrt{-g}
\mathcal{L}_{m})}{\partial g^{\mu\nu}}$. 
Using the Lagrangian \eqref{5},
the symmetric tensor $\Theta_{\mu\nu}$
finds the form
\begin{equation}
\Theta_{\mu \nu}=-\partial_{\mu}
\phi\partial_{\nu}\phi-T_{\mu\nu}
=-2\partial_{\mu}\phi\partial_{\nu}\phi-g_{\mu\nu}
\left(\frac{1}{2}\dot \phi^2 -V(\phi)\right).
\end{equation}

The effective energy-momentum tensor \eqref{24} obviously
defines the following effective energy density and pressure
\begin{equation}
\label{21}
\rho^{({\rm eff})}=\frac{1}{2}\dot \phi^2 
(1+2\beta F)+(1+4\beta F)V,
\end{equation}
\begin{equation}
\label{22}
p^{({\rm eff})} =\left[\frac{1}{2} \dot
\phi^2(1+2\beta F)-(1+4\beta F) V\right].
\end{equation}

Therefore, by replacing the FRW metric and the foregoing
energy-momentum tensor in Eq. 
\eqref{17}, the Friedmann equations
possess the same conventional form as Eqs. \eqref{2},
\begin{equation}
\label{23}
H^2=\frac{\kappa \rho^{(\rm eff)}}{3}= \frac{\kappa}{3}
\left[\frac{\dot \phi^2}{2}(1+2\beta F)
+(1+4\beta F)V\right],
\end{equation}
\begin{equation}
\frac{\ddot{a}}{a}=-\frac{\kappa}{6}\big(3p^{(\rm eff)}
+\rho^{(\rm eff)}\big)=-\frac{\kappa}{3}
\left[\dot \phi^2(1+2\beta F)-(1+4\beta F)V\right],
\end{equation}
\begin{equation}
\label{25}
{\dot H}=\frac{\ddot{a}}{a}-H^2= -\frac{\kappa}{2}
\dot\phi^2\left(1+2\beta F\right).
\end{equation}

Similarly, applying the same procedure 
which yielded Eq. \eqref{4},
we can construct a continuity
equation for $\rho^{(\rm eff)}$ and $p^{(\rm eff)}$.
Besides, by combining Eq. \eqref{25} with the time
derivative of Eq. \eqref{23} we receive the modified
Klein-Gordon equation
\begin{equation}
(1+2\beta F)[\ddot \phi+3H\dot\phi]+\beta
F_{,\phi}\dot\phi^2+(1+4\beta F)
V_{,\phi} +4\beta F_{,\phi}V=0.
\end{equation}

The slow-roll approach induces the conditions
${\dot \phi^2}\ll V$, $\vert\ddot \phi \vert \ll \vert 3H \dot \phi\vert$
and $F_{,\phi}\dot\phi^2\ll H \dot \phi$.
Thus, in these approximations, the Friedman and the modified
Klein-Gordon equations take the features
\begin{equation}
\label{70}
3H\dot\phi(1+2\beta F)+(1+4\beta F)V_{,\phi} +4\beta
F_{,\phi}V\simeq0,
\end{equation}
\begin{equation}
\label{71}
H^2\simeq\frac{\kappa}{3}(1+4\beta F)V.
\end{equation}

The potential $V(\phi)$ and the functional $F(\phi)$
give the following formulas for the slow-roll parameters
\begin{equation}
\label{30} \epsilon_V\simeq
\frac{1}{2\kappa \big(1+2\beta F\big)}
\left(\frac{V_{,\phi}}{V}+\frac {4\beta F_{,\phi}}{1+4\beta
F}\right)^2,
\end{equation}
\begin{equation}
\label{31} \eta_V\simeq\frac{1}
{\kappa\big(1+2\beta F\big)}
\left[\frac{V_{,\phi\phi}}{V}
+\frac{2\beta\big(3+4\beta F\big)
F_{,\phi}}{\big(1+2\beta F\big) \big(1+4\beta F\big)}
\frac{V_{,\phi}}{V}+\frac{4\beta
\big(1+2\beta F\big) F_{,\phi
\phi}-8\beta^2F_{,\phi}^2}
{\big(1+2\beta F\big) \big(1+4\beta
F\big)}\right].
\end{equation}

In the limit $\beta \to 0$, these parameters prominently reduce
to ${\epsilon_{\rm E}}$ and $\eta_{\rm E}$, as expected.
In fact, to fit our inflationary model predictions
with the observations the parameter
``$\beta$'' should be small. The linear
order expansions offer precise 
approximations in these conditions.

The number of the e-folds, or 
$N =\ln(a)$, is another significant
number which indicates how far the spacetime has been expanded.
For this specific generalized gravity, $N$ is given by
\begin{equation}
N\simeq\int_{\phi_i}^{\phi_f} \left[\frac{\kappa V}{V_{,\phi}}
+ \frac{2\kappa\beta V\big(FV_{,\phi}
-2F_{,\phi} V\big)}{V^2_{,\phi}}\right]
{\rm d}\phi,
\end{equation}
where the value of the inflaton at the horizon crossing is
the upper limit $\phi_f$, and $\phi_i$
is the inflaton value at the end of the inflation.

The scalar power spectrum $A_s$, which is  
associated with the curvature perturbations, along
with the scalar spectral index, which is expressed as $n_s=
1+\frac{d\ln A_s}{d\ln k}$, can be formulated 
via the slow-roll
approximation as (\textcolor{blue}{de Bruck \& Longden .2016}),
\begin{equation}
\label{60}
A_s=\frac{3\kappa H^2}{24\pi^2 \epsilon_V}\simeq
\frac{\kappa^2(1+4\beta F)V}{24\pi^2 \epsilon_V},
\end{equation}
and also
\begin{equation}
\label{61}
n_s=1+\frac{d\ln A_s}{d\ln k}\simeq 1+2\eta_V-6\epsilon_V.
\end{equation}
The Planck, BICEP/Keck, and further 
observations together provide
the following constraints on this 
amplitude (\textcolor{blue}{Galloni
et al .2023}),
\begin{equation}
A_s=(2.10\pm0.03)\times10^{-9}.
\end{equation}

In this analysis the scalar spectral index,
can be rewritten as
\begin{equation}
n_s=1+\frac{d\ln A_s}{d\ln k}\simeq 
1+\frac{d\ln A_s}{d N}=
1+\frac{d\ln A_s}{H dt}=
1+\frac{\dot\phi}{H A_s}\frac{dA_s}{d\phi},
\end{equation}
where it is recognized that \( d \ln k 
\simeq dN \) (\textcolor{blue}{Liddle
\& Lyth .2000}). Thus, by applying 
Eq. \eqref{60}, we are able to
give an alternative expression 
for $n_s$,
\begin{equation}
\label{72}
n_s=1+\frac{\dot\phi}{H}\Big[\frac{4\beta F_{,\phi}}
{1+4\beta F}+\frac{V_{,\phi}}{V}-\frac{(\epsilon_V)_{,\phi}}
{\epsilon_V}\Big],
\end{equation}
in which we employed 
Eqs. \eqref{70}, \eqref{71} and 
\eqref{30}. As a result, by applying 
Eq. \eqref{31} and utilizing
Eq. \eqref{72}, we acquire 
Eq. \eqref{61}. This approach
illustrates that, in the context 
of the tensor perturbations, generated
during the inflationary epoch, the amplitude of 
the gravitational waves finds the form 
(\textcolor{blue}{Ade et al. 2014}),
\begin{equation}
\label{80}
A_t=\frac{2\kappa}{\pi^2}H^2\simeq
\frac{2\kappa^2(1+4\beta F)V}{3\pi^2}.
\end{equation}
Thus, the relation between the 
tensor-to-scalar ratio $r = A_t/A_s$ and
the slow-roll parameter $\epsilon_V$ 
can be derived from Eqs.
\eqref{80} and \eqref{60} as follows
\begin{equation}
r\simeq16\epsilon_V.
\end{equation}
Additionally, the running of the spectral index
is defined by 
\begin{equation}
\alpha_s = \frac{{\rm d}n_s}{{\rm d}\ln\kappa}.
\end{equation}

Most often, the slow-roll 
approximation is used to compute the
valuable inflationary characteristics. 
The potential $V(\phi)$ and the
functional $F(\phi)$ define the 
slow-roll parameters. The assumption that
the slow-roll parameters are modest 
relative to unity is similar
to using the slow-roll approximation 
for the GR models. They include
the minimally coupled scalar fields 
(\textcolor{blue}{Liddle et al .1994}).
Using the cosmic evidences, we shall 
evaluate the feasibility of the
$F(\phi)T$-gravity model. Using the Planck 2018
TT(TT,TE,EE)+lowE+lensing data, the $\Lambda$CDM
model's scalar-to-tensor ratio ``$r$'', spectral
index ``$n_{s}$'', and running of scalar spectral
index ``$\alpha_{s}$'' as
(\textcolor{blue}{Akrami et al .2018;
Younesizadeh \& Younesizadeh .2021})
\begin{eqnarray}
\label{81}
r&<&0.065,
\nonumber\\
n_s&=&0.9587\pm0.0056\ (0.9625\pm0.0048),
\nonumber\\
\alpha_s&=&0.013\pm0.012\ (0.002\pm0.010).
\end{eqnarray}

%%%%%%%%%%%%%%%%%%%%%%%%%%%%%%%%%%%%%%%%%%%%%%%%%%%
\section{Calculations with the special potential and $F(\phi)$}
\label{400}

In this section, we introduce the Natural models. The
``fine-tuning'' problem of the inflation
was initially resolved by the proposal of the Natural
inflation. The inflaton potential $V(\phi)$
should be sufficiently flat to induce enough
inflation. This is a consequence of the shift
symmetries (\textcolor{blue}{Freese 
et al .1990; Galante et al .2015}).

The Natural inflation is a well-motivated
inflation model. This is because of
a quasi-shift symmetry that guarantees the
flatness of the inflationary potential,
which is required by the slow-roll approach.
Such potentials imply a suitable normalization for the
microwave background anisotropies.

The Goldstone bosons have been widely used
in the Natural models (\textcolor{blue}{Peccei \& Quinn
.(1977, 1977)}). They are produced
when a continuous symmetry is spontaneously broken.
The early inflationary expansion
of the universe has been reliably 
explained by the Natural inflation,
which is consistent with all CMB statistics. The important
discovery by the BICEP2, from the 
experiment of the CMB polarization
at the large angular scale, can be
taken as the gravitational wave signature
of the Natural inflation models.

The inflaton is an axion-like particle that can have the
potential $V\sim1-\cos\left(\frac 
{\phi}{f}\right)$, where the parameter
``$f$'' is the decay constant. 
Precisely, ``$f$'' determines
the amplitude of the inflaton 
potential in the Natural inflation
model. A larger ``$f$'' leads to a 
more flat potential, which
is consistent with the observations. 
In fact, the observational constraints on the
value of ``$f$'', in the Natural 
inflation model, are not well-defined.
It is generally understood that 
``$f$'' should be sufficiently larger
than the Planck mass to ensure a 
successful inflationary epoch.

The constraints on the CMB and 
``$r$'' can be used to estimate the
allowed range for ``$f$''. Thus, 
the range of ``$f$'' typically is
between $10^{15} - 10^{16}$ GeV. 
Depending on the employed
model, this range can vary. In fact, 
the constraints on ``$f$'' are still under
investigation and they may change.

For the natural inflation with the potential $V(\phi)$,
we suggest the following functional $F(\phi)$,
\begin{equation}
\label{4.1}
V(\phi)={V_0}  \left[1
-\cos\left(\frac{\phi}{f} \right)\right]
,\qquad F(\phi)={F_0}  \left[1
-\cos\left(\frac{\phi}{f}\right)\right].
\end{equation}

These obviously have the shift symmetry 
$\phi \to \phi + 2\pi n f$
for any integer number ``$n$''. 
Substituting these functionals
into Eqs. \eqref{30} and \eqref{31} 
allows us to derive the slow-roll
parameters $\epsilon_V$ and $\eta_V$,
\begin{eqnarray}
\label{4.2}
\epsilon_V &\simeq& \frac{
\Bigg(\frac{\sin \left( {\frac {\phi}{f}} \right)}
{h(\phi)}+\frac{4\,\beta\,\sin 
\left( {\frac {\phi}{f}} \right)
}{1+4\,\beta\, h(\phi)
}\Bigg)^2}{2 f^2 \kappa\, \left[ 
1+2\,\beta\, h(\phi) \right]},
\end{eqnarray}
\begin{eqnarray}
\label{4.3}
\eta_V&\simeq&\frac{1}{\kappa f^{2}\left[ 1+2\beta
h(\phi)\right]}\Bigg\{\frac{2\beta\left[ 3+4\beta
h(\phi)\right]\sin^{2} \left( {\frac {\phi}{f}} \right)
}{ \left[ 1+4\beta h(\phi)\right] 
\left[ 1+2\beta h(\phi)\right]h(\phi)}
\nonumber\\
&+&\frac{\cos \left( {\frac {\phi}{f}} \right)}{h(\phi) }
+\frac{4\beta \cos \left({\frac 
{\phi}{f}} \right)\left[ 1+2\beta
 h(\phi)\right]-8{\beta}^{2}\sin^{2} 
\left( {\frac {\phi}{f}} \right)
}{\left[1+4\beta h(\phi)\right]
\left[1+2\beta h(\phi)\right]}\Bigg\},
\end{eqnarray}
where $h(\phi)$ is defined as follows
\begin{equation}
h(\phi) = 1-\cos\left(\frac{\phi}{f}\right).
\end{equation}
As we said, in the limit $\beta\to 0$ these 
slow-roll parameters explicitly reduce
to $\epsilon_{\rm E}$ and $\eta_V$, as expected.
Similar to Eqs. \eqref{2.10}, the 
equations \eqref{4.2} and \eqref{4.3}
enable us to calculate the parameters 
``$r$'', ``$n_s$'' and $\alpha_s$,
\begin{eqnarray}
r=\,{\frac {8}{\kappa f^2 \left[ 
1+2\,\beta\,h \left( \phi \right) 
 \right]}\sin^2 \left(\frac {\phi}{f}\right)
\left(\frac {1}{h(\phi)}
+\,\frac {4\beta}{1+4\,\beta\,h
 \left( \phi \right)}\right)^2}
\end{eqnarray}
\begin{eqnarray}
n_{s}&=&1+\frac{1}{\kappa f^2 
\left[ 1+2\,\beta\,h \left( \phi \right)  \right] 
}\nonumber\\
&\times&\Bigg\{2\Bigg(\frac{\cos 
\left( {\frac {\phi}{f}} \right) 
}{h \left( \phi \right)} 
+\frac{2\,\beta\, \left[ 3+4\,\beta\,h 
\left( \phi \right)  \right] 
\sin^2 \left( {\frac {\phi}{f}} \right)
}{h \left( \phi \right)  \left[ 
1+4\,\beta\,h \left( \phi
\right)  \right]  \left[ 1+2\,\beta\,
h \left( \phi \right)  \right]}
\nonumber\\
&+&\frac{4\,\beta\, \left[( 1+2\,
\beta\,h \left( \phi \right)  \right] \cos
 \left( {\frac {\phi}{f}} \right) 
-8\,{\beta}^{2} \sin^2 \left( {
\frac {\phi}{f}} \right)
}{\left[( 1+4\,\beta\,h \left( \phi 
\right)  \right]  \left[( 1+2\,\beta
\,h \left( \phi \right)  \right]
}\Bigg)
\nonumber\\
&-&3\sin^2 \left(\frac {\phi}{f}\right)
\left(\frac {1}{h(\phi)}
+\,\frac {4\beta}{1+4\,\beta\,h
\left( \phi \right)}\right)^2\Bigg\},
\end{eqnarray}
\begin{eqnarray}
\alpha_{s}&=&-\frac{1}{\kappa f^2 
\left[ 1+2\,\beta\,h \left( \phi \right)  \right] 
}\nonumber\\
&\times&\Bigg\{2\Bigg(\frac{\cos 
\left( {\frac {\phi}{f}} \right) 
}{h \left( \phi \right)} 
+\frac{2\,\beta\, \left[ 3+4\,\beta\,h 
\left( \phi \right)  \right] 
\sin^2 \left( {\frac {\phi}{f}} \right)
}{h \left( \phi \right)  \left[ 
1+4\,\beta\,h \left( \phi
\right)  \right]  \left[ 1+2\,\beta\,
h \left( \phi \right)  \right]}
\nonumber\\
&+&\frac{4\,\beta\, \left[( 1+2\,
\beta\,h \left( \phi \right)  \right] \cos
\left( {\frac {\phi}{f}} \right) 
-8\,{\beta}^{2} \sin^2 \left( {
\frac {\phi}{f}} \right)
}{\left[( 1+4\,\beta\,h \left( \phi 
\right)  \right]  \left[( 1+2\,\beta
\,h \left( \phi \right)  \right]
}\Bigg)
\nonumber\\
&-&3\sin^2 \left(\frac {\phi}{f}\right)
\left(\frac {1}{h(\phi)}
+\,\frac {4\beta}{1+4\,\beta\,h
\left( \phi \right)}\right)^2\Bigg\}.
\end{eqnarray}

We should note that the 
presence of the term $F(\phi) T$ in the action 
lead to a higher level of the non-Gaussianity.
The non-Gaussianity originates from the higher-order
interactions and the non-linearities during the inflation.
This term effectively changes the scalar 
field dynamics and the curvature perturbation.
The curvature perturbation $\zeta$ can be expanded as
\begin{equation}
\zeta = \zeta_g + \frac{3}{5} f_{\rm NL}^{(\rm local)}
\zeta^2_g + \cdots .
\end{equation}
The compatibility between the parameters 
$f_{\rm NL}^{(\rm local)}$ and $n_s$ 
exhibits the following relation 
(\textcolor{blue}{Maldacena .2023}),
\begin{equation}
f_{\rm NL}^{(\rm local)}=\frac{5}{12}(1-n_s)
=\frac{5}{12}\alpha_s.
\end{equation}
Thus, the inclusion of the term $F(\phi) T$ 
in the action introduces new couplings 
and the higher-order interactions that enhance the
level of the non-Gaussianity in the cosmological perturbations.

By adjusting the values of the 
parameters ``$\beta$'' and ``$f$'',
via fitting with the data from the Planck satellite,
we can always extract appropriate 
values for ``$r$'', ``$n_s$'' and $\alpha_s$.
Besides, by selecting suitable values of
``$\beta$'' and ``$f$'', we can 
cover the entire Planck data surface.

Here, we assume a small coupling constant
``$\beta$'', and ``$f$'' as a changing variable.
Hence, we choose $\beta = 0.1$, and ``$f$'' changes between
the eleven important values. Precisely, 
we have $\beta=0.1$, $N=60 , 50$, 
and $f$ varies from 5 to 6.
Every pair $(\beta , f)$ possesses 
its own inflationary model.
The following tables represent the corresponding
values of ``$r$'' and ``$n_s$''. 
\begin{table}[!h]
\begin{center}
\caption{}
\label{demo-table}
\scriptsize{\begin{tabular}{||c c c c c c c c c c c c|}
\hline
$N$ & 60 & 60 & 60 & 60 & 60  & 60 & 60&60& 60&60 &60\\
\hline
$\beta$  &  0.1 &  0.1 &  0.1 &  0.1 &  0.1
&  0.1 &  0.1 & 0.1&  0.1 & 0.1&0.1\\
\hline
$f$ & 5 & 5.1 & 5.2 & 5.3 & 5.4 & 5.5&5.6&5.7&5.8&5.9&6 \\
\hline
$r$  & 0.0061 &0.0068  & 0.0076 & 0.0084&0.0093
&0.0101& 0.0110&0.0119&0.0127&0.0136&0.0145\\
\hline
$n_s$  & 0.9572 & 0.9586 & 0.9599 & 0.9612
&0.9623&0.9634 &0.9644&0.9653&0.9662&0.9670&0.9678 \\
\hline
\end{tabular}}
\end{center}
\end{table}

\begin{table}[!h]
\begin{center}
\caption{}
\label{demo-table}
\scriptsize{\begin{tabular}{||c c c c c c c c c c c c|}
\hline
$N$ & 50 & 50 & 50 & 50 & 50  & 50 & 50&50& 50&50 &50\\
\hline
$\beta$  &  0.1 &  0.1 &  0.1 &  0.1
&  0.1 &  0.1 &  0.1 & 0.1&  0.1 & 0.1&0.1\\
\hline
$f$ & 5 & 5.1 & 5.2 & 5.3 & 5.4 & 5.5&5.6&5.7&5.8&5.9&6 \\
\hline
$r$  & 0.0119 &0.0130  & 0.0142 & 0.0154
&0.0165&0.0177& 0.0188&0.0200&0.0211&0.0222&0.0232\\
\hline
$n_s$  & 0.9558 &0.9571 & 0.9584 & 0.9595
&0.9606&0.9616 &0.9625&0.9634&0.9642&0.9650&0.9657 \\
\hline
\end{tabular}}
\end{center}
\end{table}

These tables illustrate that the 
range of the running of the scalar 
spectral index is around 0.025. Therefore, according
to Eq. \eqref{81}, which is the result from 
the Planck 2018 data, our outcome
is consistent with the result of the Planck data.

Besides the foregoing tables, let $f$ change and assume
$\beta = 0.17$. This case has been only shown in the
figure 1 with the green color.

\begin{center}
\begin{figure}
\centering
\includegraphics[width=15cm]{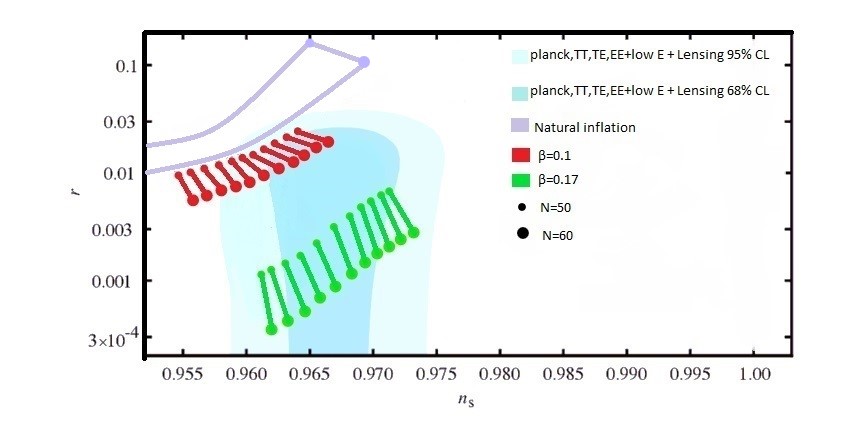}
\caption{\scriptsize{The index
$n_s$ and the ratio $r$ in the modified gravity,
anticipated by the Natural inflation 
models. As a comparison, the
gravity outcome by the Einstein 
gravity has been displayed (see the purple
band). For $\beta=0.1$ and $\beta=0.17$, 
with $5\leq f \leq 6$, see the red
and green colors, respectively.
From the (\textcolor{blue}{Planck 
Collaboration .2020}), the blue and
light blue, respectively, are the
68\% and 95\% C.L. marginal areas for $n_{s}$ and $r$ at
$k=0.002 Mpc^{-1}$.}}
\end{figure}
\end{center}

\newpage
The Natural inflation with a single-field model
in the Einstein's gravity incorporates the purple
region of the Planck/BICEP/Keck data.
By selecting appropriate values for the parameters
``$\beta$'' and ``$f$'', the term $F(\phi)\;T$ enables us to
cover the area in the $(n_s, r)$-space, favored
by the Planck 2018. Besides, by using a
special value for ``$\beta$'' and then varying
``$f$'', some parts of the Planck data has been covered.
This region has been exhibited with the red color.
Fig.1 illustrates that when $0.1 \leq \beta \leq 0.2$,
and by increasing the value of ``$f$'', we can almost cover
the whole Planck data region.
In addition, by changing the value of 
``$f$'' from 5 to higher values,
we can cover the entire region of the Planck data.

%%%%%%%%%%%%%%%%%%%%%%%%%%%%%%%%%%%%%%%%%%%%%%%%%%%%%%%%%%%%%%
\section{Conclusions}
\label{500}

The investigations via the extended gravities motivated us to
introduce the novel term $F(\phi) T$ in the gravity action.
For reducing to the Einstein's gravity,
after decaying the inflation, i.e. when $\phi =0$,
the functional $F(\phi)$ should manifestly vanish. According to
this term, we investigated the inflation theory.
At first, the inflation dynamics,
induced by an arbitrary functional $F(\phi)$ and the
potential $V(\phi)$, was studied.
Then, for special forms of $F(\phi)$ and $V(\phi)$,
we calculated the numerical values of the cosmological
observables ``$r$'' and ``$n_{s}$''. Thus,
in accordance with the experimental data, we applied the
Natural inflation model to compare its results with our
outcomes.

We applied some numerical values for 
the coupling constant ``$\beta$''.
Precisely, $\beta = 0.1$ and $0.17$ were chosen,
and the parameter ``$f$'' was left as a variable
to obtain ``$r$'' and ``$n_{s}$''. All the data have been
shown in the tables 1 and 2. After that,
we accurately compared our results
with the outcome of the standard Natural inflation.
As it is clear in the Fig.1, our results
completely agree with the Planck data.
By changing ``$\beta$'' in the range $0.1 \leq \beta \leq 0.2$
and also setting some specific values for ``$f$'',
we can recover the entire Planck data surface with higher
accuracy. Note that the case $\beta=0$ elaborated
the identical results for the standard Natural inflation.
Generally, if we select more different
values for the parameter 
``$\beta$'' from the interval [0.1, 0.2],
all the sweet-spot of the Planck 
data can be completely covered.

%%%%%%%%%%%%%%%%%%%%%%%%%%%%%%%%%%%%%%%%%%%%%%%%%%%%%%%%%%%%%%

\end{document}